\documentclass[%
 reprint,
superscriptaddress,
amsmath,amssymb,
aps,
prb
]{revtex4-1}

\usepackage{graphicx}
\usepackage{dcolumn}
\usepackage{bm}
\usepackage{epstopdf}

\begin{document}

\title{Pair Breaking Caused by Magnetic Impurities in the High-T$_\text{C}$ Superconductor Bi$_{2.1}$Sr$_{1.9}$Ca(Cu$_{1-x}$Fe$_{x}$)$_{2}$O$_{y}$}

\author{S. Parham}
\author{T.J. Reber}
\author{Y.Cao}
\author{J.A. Waugh}
\affiliation{Department of Physics, University of Colorado at Boulder, Boulder, CO 80309, USA}
\author{Z. Xu}
\author{J. Schneeloch}
\author{R.D. Zhong}
\author{G. Gu}
\affiliation{Condensed Matter Physics and Materials Science Department, Brookhaven National Laboratory, Upton, New York, 11973, USA}
\author{G. Arnold}
\author{D.S. Dessau}
\affiliation{Department of Physics, University of Colorado at Boulder, Boulder, CO 80309, USA}

\date{\today}

\begin{abstract}
Conventional superconductivity is robust against the addition of impurities unless the impurities are magnetic in which case superconductivity is quickly suppressed.  Here we present a study of the cuprate superconductor Bi$_2$Sr$_2$Ca$_1$Cu$_2$O$_{8+\delta}$ that is intentionally doped with the magnetic impurity, Fe.  Through the use of our Tomographic Density of States (TDoS) technique, we find that while the superconducting gap magnitude is essentially unaffected by the inclusion of iron, the onset of superconductivity, T$_\text{C}$, and the pair-breaking rate are strongly dependent and correlated. These findings suggest that, in the cuprates, the pair-breaking rate is critical to the determination of T$_\text{C}$ and that magnetic impurities do not disrupt the strength of pairing but rather the lifetime of the pairs.
\end{abstract}

\maketitle
Dirty superconductors are superconductors that contain significant impurities, whether they were added intentionally or not.  Conventional BCS superconductors are robust against normal impurities, but ruined with the addition of just a few magnetic impurities \cite{MattiasMagneticImpurity} like nickel or manganese.  This difference arises because the spin flip that occurs when an electron scatters off a magnetic impurity violates time reversal symmetry and thus breaks the pair, whereas simple scattering without the spin flip cannot break the pair \cite{AndersonDirty,Maki}. The importance of pair breaking scattering is clearly evident in conventional superconductors as seen by rapid the decrease in T$_\text{C}$.

Understanding the effects of impurities on the high temperature superconducting cuprates is even more critical as the cuprates are a very disordered system with many inherent defects. One of the best ways to study the effects of these defects is to intentionally add more in a controlled manner, through the addition of impurities \cite{AlloulImpurityRMP,BalatskyImpurityRMP}. There is still some debate whether impurities reduce T$_\text{C}$ by decreasing the superconducting volume \cite{NachumiZnDoped} or increasing the pair-breaking scattering \cite{BernhardZndoped}. Therefore, cuprate impurity studies require a good way to measure the effects of impurities, such as pair breaking scattering, and due to the material's anisotropy, to do so in a momentum resolved way. Previous studies on impurities in the cuprates have relied on bulk measurements \cite{LSCOimpurity,HedtInPlane} or position sensitive spectroscopies \cite{hudson2001interplay,panZnSTM}, both of which must average over momentum space and so provide less direct information about the d-wave system.

Angle Resolved Photoemission Spectroscopy (ARPES) is an excellent probe to address these challenges because it probes the band structure directly in the momentum domain. It is expected that adding impurities broadens the band by increasing scattering and this should be directly observable with ARPES. However, as we show below, the Fe impurities are only a small subset of the scattering events and so obtaining quantitative information about the Fe impurities is difficult with direct ARPES spectra. In the present paper we show, using our tomographic density of states (TDoS) technique, that we can separate the pair-breaking from non pair-breaking scattering and extract the pair-breaking induced by Fe impurities, which is $1.25$ meV/(\% Fe).

Here we present a study where we dope Bi$_{2.1}$Sr$_{1.9}$CaCu$_{2}$O$_{8+\delta}$ (Bi2212) with the magnetic impurity \cite{benseman2011valency}, Fe. A series of single crystal samples of Bi2212 with varying concentrations of Fe were grown using the floating zone method\cite{gu1994growth}. Six doping levels were studied, with Fe concentration ranging from 0\% to 2.2\%. Each sample's T$_\text{C}$ was measured using SQUID magnetometry. We chose Fe impurities because they are known to substitute for Cu \cite{gu1994growth} in the CuO$_2$ plane. Consequently, the impurity potential is poorly screened and therefore a more significant perturbation than out-of-plane impurities. Increasing Fe suppresses T$_\text{C}$ from 91K at 0\% Fe to 67K at 2.2\% Fe, a 26\% decrease over the full range of doping (inset to FIG.~\ref{fig3}). The T$_\text{C}$ suppression from Fe is stronger than that reported from either non-magnetic Zinc \cite{KanigelZnDoped} or magnetic Ni \cite{KuoNiBi2212} impurities, suggesting Fe has unique physics in this class of Cu-substituting impurities.

\begin{figure}[h]
\includegraphics[width=0.5\textwidth]{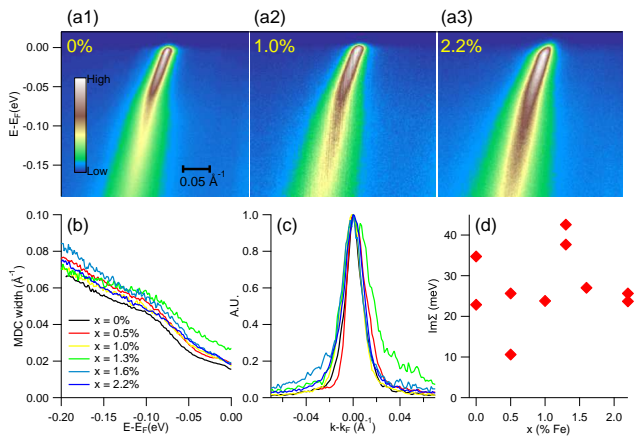}
\caption{\label{fig1}(Color online) Panels a1-a3 show raw nodal ARPES of Bi2212 across the measured Fe concentrations. Each spectrum was taken in ${\Gamma}$Y orientation at the node and at low temperature. Panel b the energy dependence of the MDC widths for all the doping levels studied.  Panel c shows the MDC’s at E$_\text{F}$ for different dopings. Panel d shows the Im $\Sigma$ extracted from the MDC at E$_\text{F}$ for every individual sample studied.}
\end{figure}

Figure~\ref{fig1} shows a compilation of the raw nodal data from various doping levels. All data were taken with 7eV photon energy and a hemispherical electron analyzer. The total experimental energy resolution was measured to be 4.5 meV using a 10K Au Fermi edge. All spectra were taken cold (T$<$30K) to minimize the effects of thermal scattering processes.  One might worry that adding too much Fe affects the crystal structure to the point where a band is unrecognizable. As shown in panels a1-a3, at all concentrations there is very clearly still a single band present, with very similar dispersions, widths, etc. Therefore, within this Fe concentration range the Fe impurities are only a weak perturbation to the electronic structure.

One of the standard methods to analyze ARPES data is momentum distribution curve (MDC) analysis \cite{Valla1999}. The MDC width is directly related to the imaginary part of the electronic self-energy, $\Sigma''$ \cite{DamascelliRMP}. However, this self-energy includes many different types of electron scattering processes, and it is hard to deconvolve the individual contributions from these different processes. In particular, the pair-breaking scattering events are a small subset of the total events, so increases in this rate may be indiscernible using the standard MDC analysis method. Indeed, as FIG.~\ref{fig1} panels b-d show, there is no clear trend in the MDC width with Fe concentration. Raw MDC’s at E$_\text{F}$, in panel c, are roughly identical and the dependence of the width with binding energy, in panel b, shows no trend beyond the sample-sample variation. Lastly, panel d summarizes these findings, showing that the Im $\Sigma(\omega=0)$, extracted from the MDC at E$_\text{F}$, has no significant trend with Fe concentration. We propose that the contributions from cleave-to-cleave variations in surface quality \cite{Valla1999} overwhelm the intrinsic changes in MDC width from the addition of Fe impurities.

To overcome the shortcomings of MDC analysis we have developed another technique to study electron interactions, the TDoS method, which is the density of states for a single slice through momentum space\cite{ReberArcs}. (Tomography is the imaging of a volume via individual slices.) The TDoS allows us to quantitatively measure both the gap magnitude and the pair-breaking scattering rate.  Briefly, to create a TDoS we isolate the coherent states of the band from the incoherent background\cite{ReberArcs}. By integrating the resulting spectrum across momentum, we obtain the coherent spectral weight. Finally, to remove effects of the Fermi distribution we divide this spectral weight by the nodal spectral weight. More details about the TDoS method can be found in Refs. \cite{ReberArcs,ReberPrePairing}.
	
\begin{figure}[h]
\includegraphics[width=0.5\textwidth]{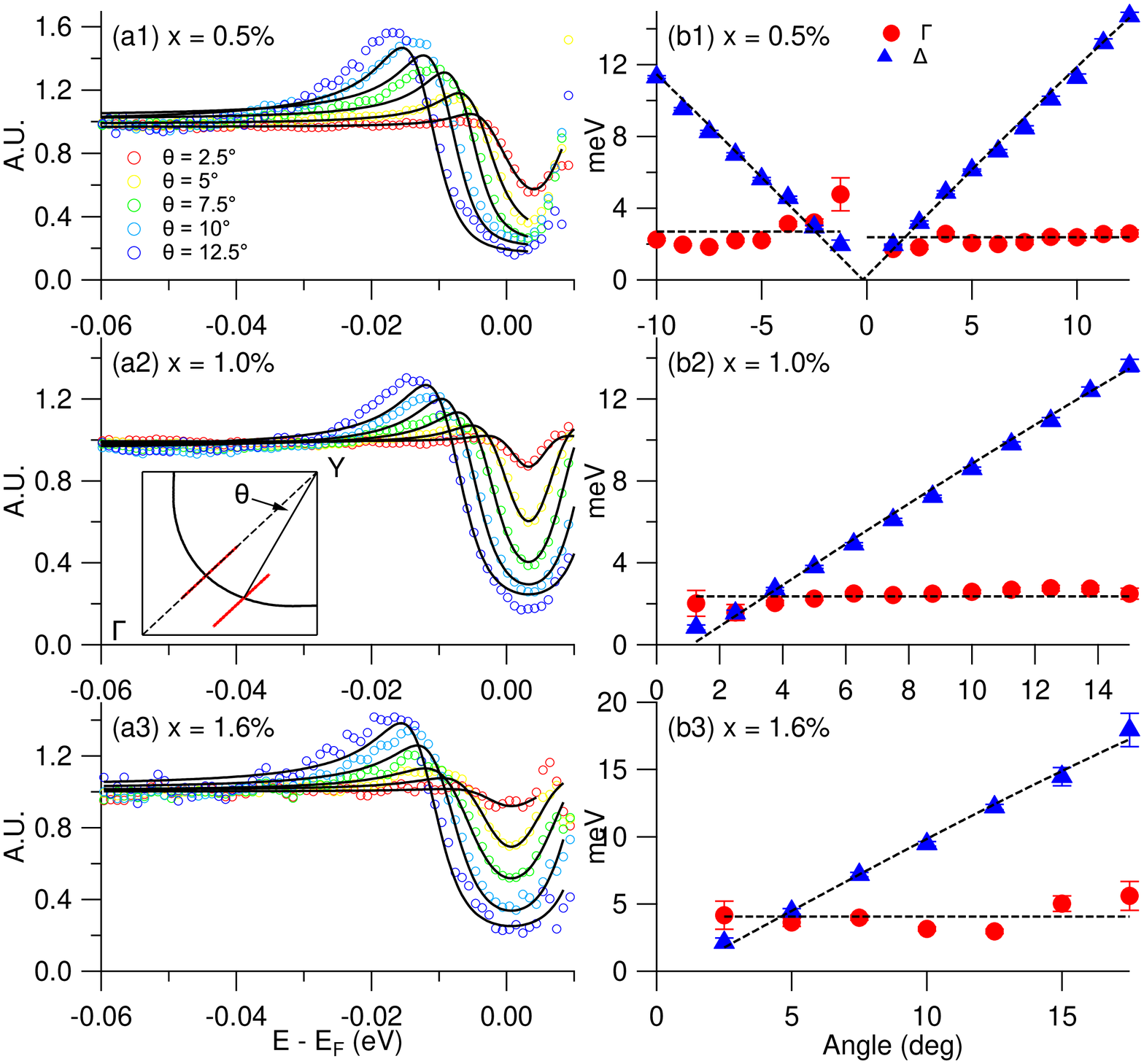}
\caption{\label{fig2}(Color online) A selection of TDoS spectra for different Fe concentrations and the corresponding TDoS fit results. Panels a1-a3 show TDoS curves (open circles) and the individual TDoS fits (black lines) for selected angles from the node. The inset in a2 shows the k-space cuts and our definition of $\theta$. Panels b1-b3 show the ${\Delta}$ and ${\Gamma_{TDoS}}$ values exacted from the TDoS fits. The fit to ${\Delta}$ is a d-wave gap with the node and gap maximum as the only fitting parameters. The fit to ${\Gamma_{TDoS}}$ is a simple average.}
\end{figure}

\begin{figure}[h]
\includegraphics[width=0.5\textwidth]{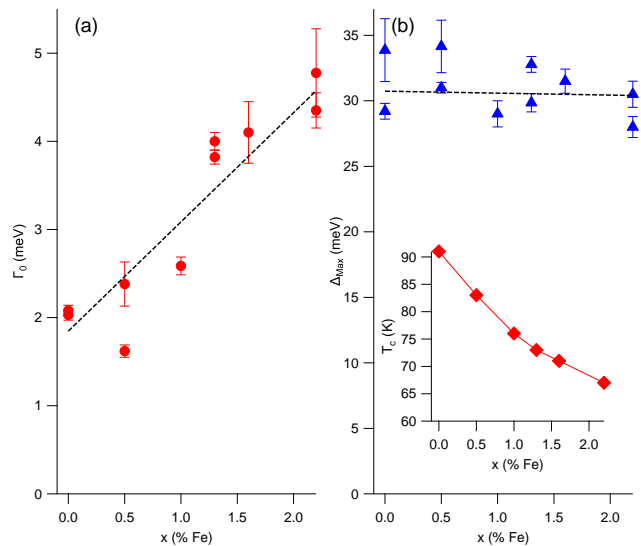}
\caption{\label{fig3}(Color online) ${\Gamma_0}$ and ${\Delta_\text{max}}$ vs. Fe impurity concentration. Panel (a) shows ${\Gamma_0}$, which is extracted from the angular average of ${\Gamma_{TDoS}}$ for a given Fe concentration. ${\Gamma_0}$ shows a linear dependence on Fe concentration, as shown by the dashed line. Panel (b) shows ${\Delta_\text{max}}$, which is extrapolated from the d-wave gap fit shown in figure 2. ${\Delta_\text{max}}$ shows very little dependence on Fe concentration, decreasing less than 1\% according to the linear fit (dashed line). The inset shows T$_{c}$ vs Fe concentration, which changes from 91K to 67K, a 26\% change over the full Fe range.}
\end{figure}

The TDoS are fit to a modified Dynes formula \cite{DynesPRL1978} that includes resolution effects. Dynes' formula, Eq.~(\ref{eq1a}), is a lifetime broadened BCS density of states. This formula was originally used in tunneling experiments on s-wave superconductors, where the gap is single valued over all of momentum space. It has been used in the cuprates in bulk transport and STM studies \cite{AlldredgeSTM,YazdaniSTM} but requires a careful integration over the d-wave gap. However, our chosen experimental geometry (inset to FIG.~\ref{fig2}a2) allows us to treat the gap as single valued. Fitting to the adapted Dynes form, shown in Eq.~(\ref{eq1b}), we extract the pairing strength, ${\Delta}$, and the pair-breaking rate, ${\Gamma_{TDoS}}$, from the TDoS spectrum.

\begin{subequations}
\begin{eqnarray}
\rho_{Dynes}(\omega)&=&Re \frac{\omega + i\Gamma_{TDoS}}{\sqrt{(\omega + i\Gamma_{TDoS})^2-\Delta^2}}\label{eq1a}
\\
I_{TDoS}(\omega)&=&\frac{\left[\rho_{Dynes}(\omega)\times f(\omega)\right]\ast R(E_\text{Res})}{f(\omega)\ast R(E_\text{Res})}\label{eq1b}
\end{eqnarray}
\end{subequations}
Here $f(\omega)$ is the Fermi function, $R(E_\text{Res})$ is an energy resolution term (4.5 meV FWHM Gaussian), and we leave the subscript TDoS on the ${\Gamma_{TDoS}}$ to distinguish it from a ${\Gamma}$ that may be determined from EDC or MDC analysis.

Earlier, we showed \cite{ReberArcs} that the TDoS method is robust against sample to sample variations in MDC widths within the same doping level. Furthermore, $\Gamma_{TDoS}$ is up to an order of magnitude smaller than the scattering rate found via MDC analysis \cite{ReberArcs}, which is consistent with $\Gamma_{TDoS}$ representing only a subset of the total scattering processes. For example, any forward scattering events should contribute to $\Sigma''$ but should not significantly increase pair-breaking and so would not be part of $\Gamma_{TDoS}$ \cite{DahmForwardScattering}. Since the MDC is more sensitive to the total scattering rate, it is also more sensitive to surface effects, such as any damage arising from cleaving, and is therefore a less intrinsic measure of the material. As we show in FIG.~\ref{fig3}, the TDoS method can extract the pair-breaking dependence on Fe concentration that is simply masked in the MDC analysis.

Figure~\ref{fig2} shows a compilation of the TDoS data for selected dopings, along with the extracted $\Delta$ and $\Gamma_{TDoS}$ values. Panels a1-a3 show the angular dependence of the TDoS along with the Dynes’ fits for three different Fe concentrations. Note how the pile-up of states increases in size and location with $\theta$, signifying an increase in the gap size. Note also that as $\Gamma_{TDoS}$ increases more pairs are broken and more weight is present at E$_\text{F}$. This trend can be seen in the $\theta=2.5^\circ$ curves (red online) moving from panel a1 to a3. One can see good quantitative agreement between the data and the fits with perhaps slightly degraded agreement at the high angle data. We attribute the disagreement to a higher order effect that is not captured by the simple nature of the functional form (which is only a two parameter fit). Shown in panels b1-b3 are the $\Delta$ and $\Gamma_{TDoS}$ values for all angles at the selected Fe concentration as well as the fit to the angular dependence. $\Delta$ rises linearly away from the node, as a d-wave gap in the near nodal regime. We extrapolate this near nodal data to obtain the maximum gap, $\Delta_\text{max}$, by fitting to a d-wave form: $\Delta(\theta)=\Delta_\text{max}|\sin(2\theta)|$. In contrast to $\Delta$, $\Gamma_{TDoS}$ is roughly constant across all angles, which is consistent with our previous TDoS study on BSCCO and is applicable over a wide doping range \cite{ReberArcs,ReberPrePairing}, but differs markedly from the result using MDC or EDC analysis \cite{Valla1999}. Due to the angular invariance, we average $\Gamma_{TDoS}$ over all angles to extract $\Gamma_0$.

Figure~\ref{fig3} shows both $\Delta_\text{max}$ and $\Gamma_0$ vs. Fe concentration. $\Gamma_0$ rises by about 130\% over the full experimental range while $\Delta_\text{max}$ remains relatively unchanged. More specifically, $\Gamma_0$ shows a linear increase with Fe concentration, which we discuss in greater detail below. Despite a slight downward slope visually, the $\Delta_\text{max}$ data show no real trend with added Fe; the uncertainty in the slope of a weighted linear fit is several times larger than the slope's value. Therefore, we conclude that there is no significant change in the superconducting gap from the addition of Fe impurities. A related experiment on Zn-substituted Bi2212 found, using the standard EDC method, that the antinodal gap size was insensitive to the addition of Zn impurities \cite{KanigelZnDoped}. In addition, Ref.~12 found a qualitative increase in the number of in-gap states  with the addition of Zn impurities, consistent with an increased $\Gamma_{TDoS}$, though the TDoS technique is more quantitative than EDC analysis. Both of these results from Ref.~12 are qualitatively similar to those reported here and suggest that the physics involved in both Zn and Fe impurities is similar.  Note also that while $\Delta_\text{max}$ didn't change, T$_\text{C}$ decreased by 26\%, indicating a departure from BCS superconductivity where these quantities are directly proportional. In addition, the fact that $\Delta_\text{max}$ is essentially unchanged but both T$_\text{C}$ and $\Gamma_0$ change indicates that pair-breaking processes are critical in setting T$_\text{C}$ in this material.

We propose a simple model based on ballistic transport that captures the essential physics of the linear increase of $\Gamma_0$ with Fe concentration. The dimensionality of $\Gamma_0$ suggests the simplest form would be a velocity divided by a length. The natural choices are the Fermi velocity and the mean distance between Fe impurities, respectively. Here we use the measured, nodal Fermi velocity, $1.9$ eV$\cdot$\AA, instead of the bare velocity because the impurities are a perturbation on a system that already has self-energy effects. In this simple model, we estimate the mean distance as the Cu-Cu spacing, $3.82$ \AA, divided by the concentration, $x$, of Fe scattering sites. And we add a dimensionless parameter, $\alpha$, that represents the 2D scattering cross-section for these events. Note that in two dimensions, the scattering cross-section has dimensions of length instead of the normal dimensions of area for a 3D scattering process. Since we have divided out the length dimension using ``a'', $\alpha$ is dimensionless and has units of Cu-Cu spacing. This model would be sufficient if Fe impurities were the only things causing scattering. However, Bi2212 has a number of native defects including Cu vacancies and several types of oxygen vacancies and dislocations \cite{EisakiDisorder,mcelroy2005stm,hudson2003stm,hoffmanthesis,zeljkovic2011stm}. Each of these scattering species should contribute to the scattering rate in the same way as outlined for Fe impurities, and so we can add a sum of terms to the formula for $\Gamma_0$, as shown in Eq.~\ref{eq2a}. This sum includes all native defects in the material, including any not listed above. However, it is beyond the scope of this paper to determine both the concentrations and cross-sections for all these defects. And since they are not relevant to extracting properties of the Fe impurities, we group all of the native defects together in a term called $\Gamma_0^\text{Native}$, shown in Eq.~\ref{eq2b}. This model should hold as long as the impurities are well separated and therefore not interacting with each other.

\begin{subequations}
\begin{eqnarray}
\Gamma_0\approx\frac{v}{l_\text{Fe-Fe}} ; l_\text{Fe-Fe}&=&\frac{a}{x} \rightarrow \Gamma_0^{Fe}(x)=\frac{v}{a}(\alpha x)\nonumber
\\
\Rightarrow\Gamma_0 &=&\frac{v}{a}(\alpha x+\sum\limits_i^N\beta_i x_i)\label{eq2a}
\\
\Rightarrow\Gamma_0 &=&\frac{v}{a}(\alpha x)+\Gamma_0^\text{Native}\label{eq2b}
\end{eqnarray}
\end{subequations}

\begin{figure}[h]
\includegraphics[width=0.5\textwidth]{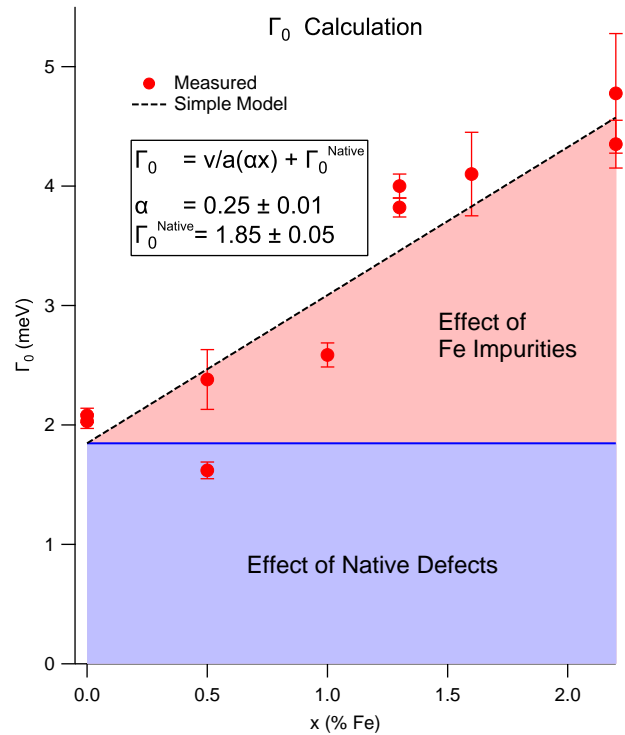}
\caption{\label{fig5}(Color online) Segregating the effects of Fe impurities from native defects using our simple model of impurity scattering. ${\Gamma_0^\text{Native}}$ is the pair breaking rate for all the combined native defects not including any Fe impurities, while $\alpha$ is a dimensionless parameter that represents the pair-breaking scattering cross-section (in units of Cu-Cu spacing) for Fe impurities. }
\end{figure}

Figure~\ref{fig5} shows the linear fit to our data in the context of this model. The linear increase in $\Gamma_0$ we attribute to Fe impurity scattering while the offset is due to the combined effects of all native defects in the material. From this fit we extract $\alpha$ = 0.25, which means the pair-breaking scattering cross-section of a single Fe impurity is approximately 25\% of the Cu-Cu spacing, or 0.96 \AA. This is a reasonable value of the cross-section for a single site defect, and suggests that Fe impurities are relatively weak pair-breakers. This result is also consistent with STM work on Fe-substituted BSCCO that claims Fe is a weaker scatterer than Ni\cite{KoopmanSTM_FeBSCCO,hudson2001interplay}, Zn\cite{panZnSTM}, and Cu vacancies \cite{hudson2003stm}, \cite{[{Ref. 29 directly compares Fe to Ni impurities. Refs. 10,11, and 26 measure the individual impurities, and a calculation of the relative strengths of Ni, Zn, and Cu vacancies is provided by }] wang2005impurity}. To our knowledge, this is the first time the pair-breaking cross-section has been measured for any impurity in the cuprates. Furthermore, this general experimental procedure would work on any impurity, as long as there is sufficient control to increase the concentration of a known impurity while keeping all others constant.

Using the TDoS technique we have measured the effects of Fe impurities on the superconducting gap and pair-breaking scattering rate in the high T$_\text{C}$ superconductor Bi2212. The main effect of Fe impurities is to increase the pair-breaking scattering rate, while leaving the gap parameter unchanged. This indicates that magnetic impurities do not affect the pairing strength, just the pair lifetimes. This result also confirms that cuprate superconductivity is sensitive to pair-breaking scattering. The correlation between $\Gamma_0$ and T$_\text{C}$ is suggestive and motivates further studies at temperatures near T$_\text{C}$, as well as studies of other impurities, such as Zn and Ni, in hopes of discovering any universal behavior of cuprate impurities.

The authors would like to thank M. Hermele, E. Calleja, and K. McElroy for helpful discussions. This research is supported by DOE grant No. DE-FG02-03ER46066 (Colorado) and by DE-AC02-98CH10886 (Brookhaven). SP is supported by the Department of Energy Office of Science Graduate Fellowship Program (DOE SCGF), made possible in part by the American Recovery and Reinvestment Act of 2009, administered by ORISE-ORAU under contract no. DE-AC05-06OR23100. The Stanford Synchrotron Radiation Laboratory is funded by the Department of Energy , Office of Basic Energy Sciences.

\bibliography{FeBSCCOrefs}

\end{document}